\documentclass[10pt]{article}

\usepackage{amsmath}
\usepackage{times}
\usepackage{bm}
\usepackage{natbib}
\usepackage{amsfonts}

\usepackage{enumerate}
\usepackage{mathscinet}

\usepackage[plain,noend]{algorithm2e}

\usepackage{graphicx}
\usepackage[margin=1cm, textfont=small]{caption}

\usepackage[textwidth=1.5cm, disable]{todonotes}
\usepackage{url}

\usepackage{booktabs,caption}
\usepackage[flushleft]{threeparttable}
\usepackage{setspace}

\allowdisplaybreaks

\newcommand{\MSE}{\text{{\sc mse}}}

\newtheorem{definition}{Definition}

\begin{document}

\title{\vspace{-1cm} Seeded intervals and noise level estimation in change point detection: A discussion of~\cite{Fryzlewicz_WBS2}}

\author{Solt Kov\'acs${}^{1}$, Housen Li${}^{2}$, Peter B\"uhlmann${}^{1}$\\
\vspace{0.1cm}\\
{\small${}^{1}$Seminar for Statistics, ETH Zurich, Switzerland}\\
{\small${}^{2}$Institute for Mathematical Stochastics, University of G\"ottingen, Germany}}

\date{June 2020}

\maketitle

\begin{abstract}
\noindent
In this discussion, we compare the choice of seeded intervals and that of random intervals for change point segmentation from practical, statistical and computational perspectives. Furthermore, we investigate a novel estimator of the noise level, which improves many existing model selection procedures (including the steepest drop to low levels), particularly for challenging frequent change point scenarios with low signal-to-noise ratios.
\end{abstract}


\vspace{0.3cm}
\paragraph{Starting remarks}
We congratulate Piotr Fryzlewicz for his interesting and stimulating paper entitled ``Detecting possibly frequent change-points: Wild Binary Segmentation 2 and steepest-drop model selection'' (see \citealp{Fryzlewicz_WBS2}) and we also thank the Editor of the Journal of the Korean Statistical Society for the invitation to write a discussion! We also take the opportunity to express that we have been very much inspired by \citeauthor{Fryzlewicz_WBS}' previous work on change point detection. We mention in particular his path-breaking work on wild binary segmentation \citep{Fryzlewicz_WBS} which has heavily shaped the thinking on change point detection as well as the narrowest over threshold method \citep{Baranowski}. Our remarks and comments should be seen in this light, owing credit to many pioneering ideas from \citeauthor{Fryzlewicz_WBS}. His current paper proposes a new change point detection algorithm (WBS2) that might be thought of as a hybrid between binary segmentation (BS, \citealp{Vostrikova}) and wild binary segmentation (WBS, \citealp{Fryzlewicz_WBS}), as well as a novel model selection procedure (steepest drop to low levels, SDLL). While recently several algorithms have been proposed for change point detection, some of which are computationally very efficient, less attention is paid to model selection which in our point of view remains a difficult task. We thus welcome and appreciate new methods such as SDLL that contribute to model selection. In the following we would like to point to a number of alternatives and modifications that might lead to improved performance in terms of stability/reproducibility, estimation error, range of extendability and computational speed.

\vspace{-0.3cm}
\paragraph{A summary of seeded binary segmentation (SeedBS, \citealp{SeedBS})} The computational cost for evaluating the CUSUM statistics in $M$ random intervals within WBS is of order $O(MT)$, with the worst case $M=O(T^2)$ (for very short minimal segment length as e.g.~in frequent change point scenarios or when aiming to obtain the ``complete solution path'' defined by \citealp{Fryzlewicz_WBS2}). In \cite{SeedBS}, we proposed a deterministic construction of ``seeded'' intervals that have a total length $O(T\log(T))$ at most and thus allow fast evaluation of the test statistics in all scenarios (as opposed to the worst case $O(T^3)$ for the random intervals in WBS). The idea is roughly to start with the full range of observations in the first layer and sequentially split the intervals from the previous layer into a left, right and an overlapping middle interval according to a chosen ``decay'' parameter (e.g.~$(1,T)$, $(1,T/2)$, $(T/4,3T/4)$, $(T/2,T)$, $(1,T/4)$, $(T/8, 3T/8)$, etc.) and stopping at the layer where the length of intervals drops to a pre-defined minimal length $m$. Seeded binary segmentation (SeedBS) with greedy selection resembles WBS, while SeedBS with the narrowest over threshold (NOT) selection is similar to the NOT method of \cite{Baranowski}, the difference in both cases is essentially to use seeded intervals instead of random ones. Seeded intervals consist of $O(T)$ intervals efficiently covering various scales, in particular guaranteeing good coverage of each single change point which in turn ensures good estimation performance (e.g.~minimax optimality for the Gaussian change in mean setup for SeedBS with NOT selection for a suitable threshold). For further details on theory, performance in simulations, explanations on where the computational inefficiency of random intervals comes from and how seeded intervals overcome this, see \cite{SeedBS}.
 
\vspace{-0.3cm}
\paragraph{Adapting SeedBS to frequent change point scenarios} In WBS2 (with $\tilde M = 100$) once the current search interval $(s,e)$ has less than $15$ observations, all intervals within $(s,e)$ are evaluated, facilitating estimation (and model selection) in frequent change point scenarios when the true segments are extremely short. A possible adaption to SeedBS in order to match this is to include all intervals below a certain length additionally to the regularly generated seeded intervals. In the experiments further below we added all intervals having less than $10$ observations. Alternatively, when aiming to improve results in very challenging scenarios, then instead of the default decay parameter of $2^{1/2}$, one could generate more seeded intervals by changing the decay to e.g.~$2^{1/8}$, resulting in an increase of computational time by a factor of approximately four. Another option for improving estimation is an adaptive version (i.e.~ASeedBS) of SeedBS that we discuss further below.

\begin{figure}[h]
\vspace*{-7pt}
\includegraphics[width=1\textwidth]{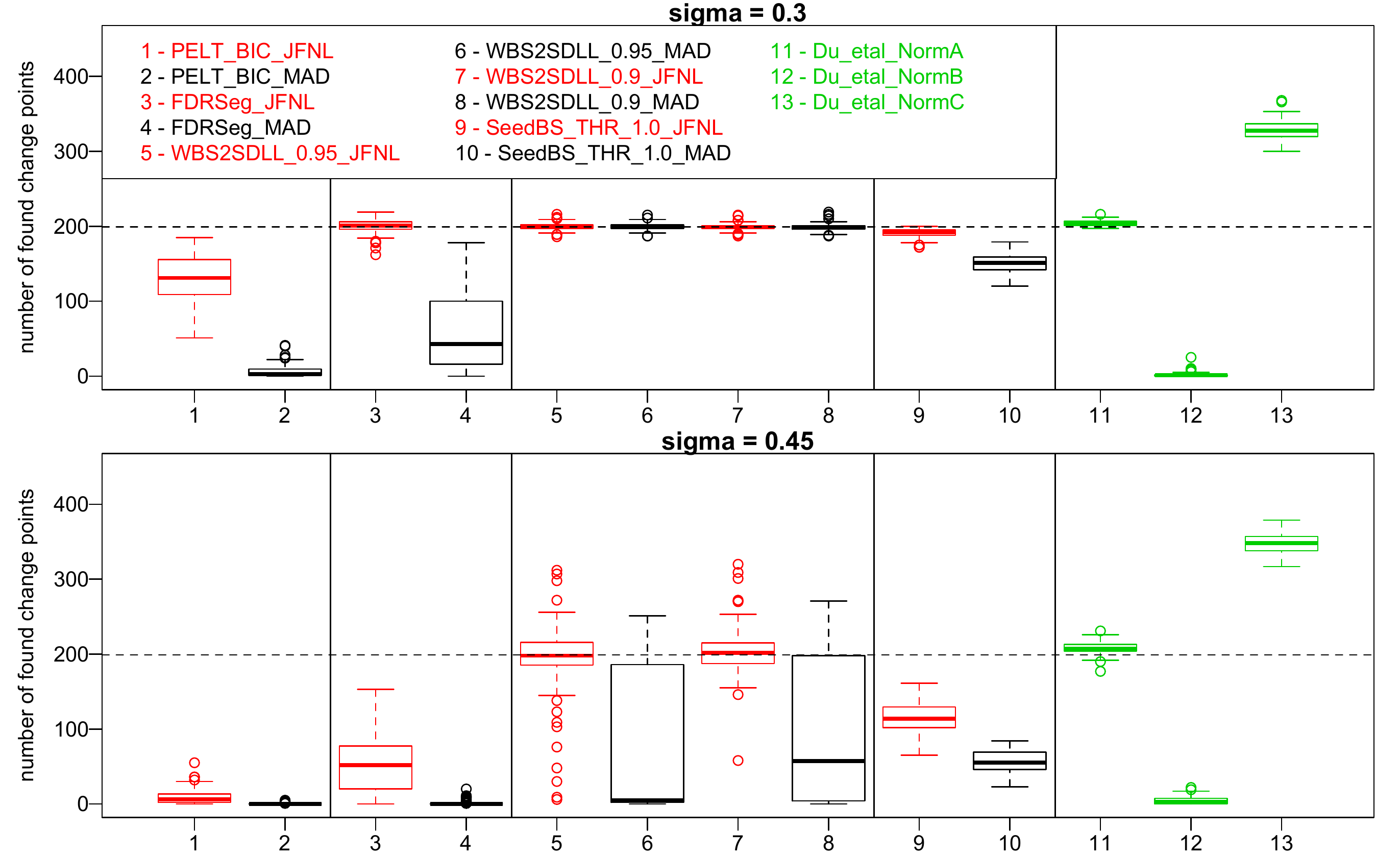}
\caption{Boxplots of the number of found change points (based on $100$ simulations each) in the \texttt{extreme.teeth} example of \cite{Fryzlewicz_WBS2} with noise level $\sigma=0.3$ (top) and $\sigma = 0.45$ (bottom) for various estimation approaches. Model selection approaches based on the JFNL estimator are in red, MAD based counterparts in black, while green refers to approaches from \cite{Du2016}. The true number of change points is indicated by the dashed horizontal line at $199$ and the vertical lines group the methods as a visual aid.}
\label{methods_with_JFN_vs_MAD}
\vspace*{-10pt}
\end{figure}

\vspace{-0.3cm}
\paragraph{Some advantages of SeedBS: a fast, stable/replicable and easy to generalize alternative}
First of all, when the last layer with the smallest considered scale within seeded intervals includes all intervals of length $m=2$ observations, one obtains a complete solution path. Hence, unlike for WBS, model selection in frequent change point scenarios is still possible with a fixed threshold. In Figure~\ref{methods_with_JFN_vs_MAD} estimators number $9$ and $10$ (``SeedBS\_THR\_1.0\_...'') use the threshold $1.0\cdot \hat\sigma\sqrt{2\log(T)}$ with the JFNL estimator (from Definition~\ref{def:noise_estimator} further below) and the MAD based estimator as $\hat \sigma$, respectively. When paired with the JFNL estimator, this fixed thresholding performs nearly as good as the best competitors (e.g.~SDLL) in the lower noise level $\sigma = 0.3$. For $\sigma = 0.45$ the best performing SDLL method performs better in general in estimating the number of change points than the fixed threshold (which is not surprising as the fixed threshold is not adapted to the data). However, as an advantage, the fixed threshold has smaller variance than the SDLL based model selection. Second, SeedBS can be combined with many model selection approaches, in particular also with the SDLL model selection. The advantage here is that the deterministic scheme utilized in the seeded intervals leads to reproducible results unlike WBS2+SDLL, where the estimated number of change points for various runs on the same data can show high variability/instability for high noise levels due to the randomness of the intervals. Aggregation over multiple WBS2 runs and taking the ``median run'' as proposed by \cite{Fryzlewicz_WBS2} might help to gain more stability, but does not fully eliminate the issue while increasing computational efforts. Third, a SeedBS-BS hybrid can be considered (similar to WBS2 being a hybrid between BS and WBS) by generating new seeded intervals specifically in between previously found change points and thus adapted to the information on the signal obtained in earlier iterations of the procedure. This adaptive seeded binary segmentation (ASeedBS) might perform slightly better than SeedBS in challenging scenarios (at the price of somewhat increased computational times as a new set of seeded intervals are evaluated each time a change point candidate is added). One advantage of ASeedBS compared to WBS2 is reproducibility due to the lack of randomness. Fourth, we would like to mention applicability in models beyond the Gaussian change in mean setup. WBS2 has a greedy nature when selecting the next best split point. This might be limiting when trying to apply WBS2 e.g.~in piecewise linear models, where the best single split point over an interval containing multiple change points may not correspond to any of the true change points (see \citealp{Baranowski} and their proposal of the NOT selection instead of a greedy one to solve this issue). In SeedBS and also ASeedBS greedy selection is not essential and they could be easily combined with other selection rules, e.g.~the NOT selection. Note that evaluating the CUSUM statistic for $\tilde M = O(\log T)$ random intervals has the same cost $O(T\log T)$ as evaluating all the $O(T)$ seeded intervals. However, if the spacings between change points are short, it is likely that none of the $\tilde M = O(\log T)$ random intervals contain only a single change point as opposed to seeded intervals, for which it is likely that some of the $O(T)$ intervals will be suitable (i.e.~contain a single change point). As an illustration of wide applicability, in \cite{SeedBS} we used SeedBS to detect changes in (high-dimensional) Gaussian graphical models using a graphical lasso based estimator as in \cite{Londschien}. Last, but not least, as a method of $O(T\log(T))$ worst case computational complexity, SeedBS is fast, in particular in our recent Fortran based implementation (planned to be made publicly available) that needed (using the default options) around $0.001$ seconds for the \texttt{extreme.teeth} signal of \cite{Fryzlewicz_WBS2} and thus about one to two orders of magnitude less than the available R implementation of WBS2. We currently work on the so-called optimistic search technique \citep{OBS} reducing the computational times for various methods. Applying this search for SeedBS results in the optimistic seeded binary segmentation (OSeedBS) method, which further reduces the computational complexity to $O(T)$ in the Gaussian change in mean setup when using the NOT selection. From a practical perspective SeedBS and OSeedBS naturally offer the possibility to parallelize the evaluation of the test statistic in various seeded intervals, which can be an advantage when applying it in more expensive high-dimensional setups such as the previously mentioned setup of \cite{Londschien}. We tried SeedBS and ASeedBS together with the SDLL model selection on the \texttt{extreme.teeth} signal of \cite{Fryzlewicz_WBS2} and the estimation performance for the number of change points was similar to WBS2+SDLL. Much bigger differences seemed to result from a different choice of noise level estimator and thus we present some simulation results for that in the next paragraph. Overall, in our point of view the main possible advantages of SeedBS compared to WBS and WBS2 are the reproducibility, easy extendability beyond the Gaussian change in mean case trough e.g.~the NOT selection and, to some extent, the easier derivation of theoretical results.

\begin{figure}[h]
\vspace*{-7pt}
\includegraphics[width=1\textwidth]{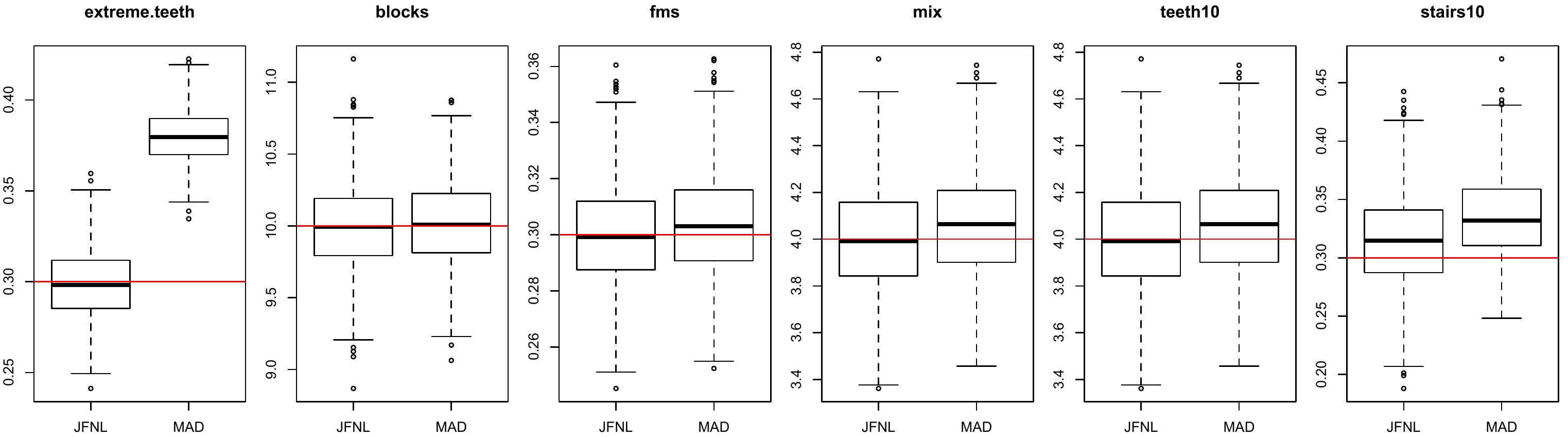}
\caption{Boxplots of the JFNL and MAD estimators of the noise level $\sigma$ in the \texttt{extreme.teeth} example of \cite{Fryzlewicz_WBS2} on the very left panel and five other examples from \cite{Fryzlewicz_WBS} based on $1000$ simulations each. The true values of $\sigma$ are indicated by the red horizontal lines respectively.}
\label{variance_estimators}
\vspace*{-10pt}
\end{figure}

\vspace{-0.3cm}
\paragraph{An alternative estimator for the error variance} A commonly used approach for the estimation of noise levels in the Gaussian change in mean setup (with constant variance) is to use a robust scale estimator (e.g.~median absolute deviation, MAD) of consecutive differences $\left({X_2-X_1}, \ldots, {X_T-X_{T-1}}\right)/{\sqrt2}$ so that jumps in the mean of the signal can be treated as outliers. However, in frequent change point scenarios, there can be many of these outliers, potentially leading to severe upwards bias. While \citet{Fryzlewicz_WBS2} wrote that accurate estimation of the noise variance ``can be difficult to achieve in frequent change point models'', we would like to show that the task is not impossible if a different strategy is considered. Rather than treating jumps as outliers, we actively consider them in the following sense: when looking at consecutive differences (with lag one), each jump in the signal is counted once, however, when looking at lag two differences, each jump is counted twice. For example, assuming there is a jump at the $k$-th observation, out of the four pairs
$(X_{k-1}-X_{k-2}, X_{k}-X_{k-1}, X_{k+1}-X_{k}, X_{k+2}-X_{k+1})$ only $X_{k}-X_{k-1}$ includes the jump, but out of four lag two differences $(X_{k-1}-X_{k-3}, X_{k}-X_{k-2}, X_{k+1}-X_{k-1}, X_{k+2}-X_{k})$ two of them, namely $X_{k}-X_{k-2}$ and $X_{k+1}-X_{k-1}$ include the jump (while the noise is of course contained additionally in both series). The assumption for this ``counting'' to work is to have minimal segment length of least 2 observations. Overall, the difference in the variance of the lag one and lag two series can be utilized to filter out the jumps as follows. 
\begin{definition}
\label{def:noise_estimator}
The jump filtered noise level (JFNL) estimator for the variance $\sigma^2$ is defined as 
\begin{equation*}
\hat\sigma^2_{JFNL} = \max\left\{0, 
2\cdot v\left(\frac{X_2-X_1}{\sqrt2}, \ldots, \frac{X_T-X_{T-1}}{\sqrt2}\right) 
- v\left(\frac{X_3-X_1}{\sqrt2}, \ldots, \frac{X_T-X_{T-2}}{\sqrt2}\right)\right\}, 
\end{equation*}
where $v(Y_1,\ldots, Y_l) =  \sum_{i=1}^l (Y_i - \bar Y)^2/l$ is the empirical variance of $Y_1,\ldots, Y_l$.
\end{definition}
Although difference based variance estimation methods have a long history in nonparametric regression (see e.g.~\citealp{HKT90}), we are not aware of JFNL having been proposed elsewhere. The goal when using the empirical variance is to actively count each of the jumps (which is the core idea when comparing the two lags), whereas robust estimators would treat a part of the jumps as outliers, but possibly to a different extent in the lag one versus the lag two differences such that their comparison becomes difficult. If the true noise level $\sigma^2$ is very low compared to the jumps, the difference of the variance terms above can become negative. In such cases the maximum in the definition above kicks in and we threshold at zero in order to have a non-negative variance estimator. While well-defined, estimating zero as a noise level is not particularly useful when applying for model selection purposes. Note however that this typically occurs only in scenarios where change point detection and model selection is an easy task with many available methods, namely in scenarios with low noise levels compared to the sizes of the jumps. In general, lag $k$ differences count jumps $k$ times and thus, potentially even better estimators could be constructed similarly if it is reasonable to assume that each segment contains at least e.g.~$k=3$ or $k=4$ observations. One could come up with other estimators as well and also try aggregating various estimators for the variance (e.g.~via the median) to have an ensemble type estimator which may even outperform individual ones. A related, but slightly different avenue is to try to make the estimator robust. In Figure~\ref{variance_estimators} we compare JFNL to the MAD estimator (of lag one differences) for various commonly used test signals. In the \texttt{extreme.teeth} signal $\hat\sigma^2_{JFNL}$ performs much better than the MAD estimator and also on the other test signals it seems to perform somewhat better. When looking at scenarios with lower noise levels, a higher variance of the JFNL estimator compared to MAD would be visible such that some other estimators could be better suited than JFNL. However, in the shown scenarios with comparably high noise (which are the difficult scenarios from a model selection perspective), JFNL seems to offer some advantages. The performance of JFNL for the ``stairs10'' signal can be further improved by taking $\tilde v(Y_1,\ldots, Y_l) =  \sum_{i=1}^l Y_i^2/l$ instead of $v$ in Definition~\ref{def:noise_estimator}. As mentioned before, it could be promising to combine strengths of an ensemble of various estimators. There is definitely room for further improvements. It remains to be tested how well $\hat\sigma^2_{JFNL}$ performs in examples beyond the ones shown here as well as in real data where the independence and minimal segment length assumptions may not hold.

\vspace{-0.3cm}
\paragraph{Utilizing the jump filtered noise level estimator $\hat\sigma^2_{JFNL}$} For methods/model selection procedures, that rely on an estimate of the noise level one can hope to obtain better results when using a better estimate of the noise level. We illustrate this for some approaches for which changing the default noise level estimate is easy to do within the corresponding R package/implementation. In Figure~\ref{methods_with_JFN_vs_MAD} we consider the \texttt{extreme.teeth} signal with noise level $\sigma = 0.3$ (top) and $\sigma = 0.45$ (bottom). As expected, the JFNL based estimators in general result in more estimated change points than the MAD variants as JFNL does not overestimate the variance. Interestingly, for the lower noise level, FDRSeg \citep{LMS16} and SeedBS \citep{SeedBS} with a fixed threshold (with JFNL) perform similar to SDLL (with WBS2 or even in combination with SeedBS that we do not show to save space). For higher noise levels, all methods shown in red (JFNL) and black (MAD) tend to find fewer change points. The exception is SDLL with JFNL used for the threshold to determine the ``low level'' of the drop, which has median estimated number of change points still close to the true number. Hence, a bit surprisingly, it seems like JFNL also improves SDLL methodology in noisy scenarios. SeedBS with a fixed threshold finds fewer change points, but with a smaller variance than SDLL. As the noise level is quite high in this scenario, the question is also what one is aiming for: a good estimate of the number of change points, potentially at the price of including more false positives, or selecting fewer change points and thus also fewer false positives? Note that we did not change the calibration/constants within SDLL when using the JFNL estimator instead of MAD, which might slightly influence the results. Also, besides the estimated number of change points, other performance measures might be helpful and necessary to understand the detailed behavior of individual methods. Lastly, the green parts of Figure~\ref{methods_with_JFN_vs_MAD} show estimators from \cite{Du2016} implemented in the \texttt{StepSignalMargiLike} package (available from Samuel Kou's homepage) with the therein proposed three priors (called NormA, NormB and NormC). With the implicit penalization, i.e. default options without us giving any estimate for the noise level, the NormA approach gave very good estimates for the number of change points (with remarkably low variance) even for $\sigma=0.45$ and thus we considered it worthwhile mentioning. The NormB estimator returned way too few and the NormC way too many candidate change points. In practice, it is a question which prior to choose, moreover, the implementation does not scale well to long time series (around 8 seconds for a time series of length $10^5$ and 80 seconds for a time series that is three times longer).

\vspace{-0.3cm}
\paragraph{BIC based model selection with $\sigma$ unknown or known?}
We note that somewhat confusingly, BIC type criteria are used differently in different publications and software for change point detection. \cite{Yao1988} and \cite{Fryzlewicz_WBS}, for example,  minimize $T/2 \cdot \log(\MSE) + \mathrm{penalty}$ (assumption of unknown variance), while dynamic programming approaches (e.g.~PELT from \citealp{Killick_etal}) typically minimize $T\cdot \MSE+\mathrm{penalty}$, with $\MSE$ denoting the mean squared errors. In the latter approach the penalty term should include the noise level which in this case is assumed to be known. Yet, in practice often an estimate $\hat\sigma$ is used in the absence of exact knowledge on $\sigma$ for which the JFNL estimator could be useful. However, such an approach ignores the uncertainty of the estimated noise level, an issue which is mentioned by \cite{Fearnhead_Rigaill} among many other interesting considerations for practice.

\vspace{-0.3cm}
\paragraph{Conclusions}
We pointed out similarities between SeedBS and WBS2, both being computationally fast methods. The deterministic construction of intervals of SeedBS helps to make results reproducible compared to the randomness inherent in WBS2. Both approaches are generic and thus suitable for many change point detection problems. Our perhaps biased and slight preference for SeedBS also includes that it seems easier to combine seeded intervals with other model selection procedures, such as the NOT method, that are essential in certain change point detection problems beyond the Gaussian change in mean case. We also introduced the JFNL estimator for the noise level and showed in simulations that using JFNL could improve SDLL based model selection in rather noisy scenarios, as well as more traditional model selection approaches relying on a direct estimate of the noise level.

\vspace{-0.3cm}
\paragraph{Acknowledgement}
Solt Kov\'acs and Peter B\"uhlmann have received funding from the European Research Council (ERC) under the European Union's Horizon 2020 research and innovation programme (Grant agreement No. 786461 CausalStats - ERC-2017-ADG). Housen Li gratefully acknowledges the support of the  Deutsche Forschungsgemeinschaft (DFG, German Research Foundation) under Germany’s Excellence Strategy - EXC 2067/1-390729940.

\bibliographystyle{apalike}
\bibliography{ref}

\begin{thebibliography}{}

\bibitem[Baranowski et~al., 2019]{Baranowski}
Baranowski, R., Chen, Y., and Fryzlewicz, P. (2019).
\newblock Narrowest-over-threshold detection of multiple change points and
  change-point-like features.
\newblock {\em Journal of the Royal Statistical Society, Series B},
  81(3):649--672.

\bibitem[Du et~al., 2016]{Du2016}
Du, C., Kao, C.-L.~M., and Kou, S.~C. (2016).
\newblock Stepwise signal extraction via marginal likelihood.
\newblock {\em Journal of the American Statistical Association},
  111(513):314--330.

\bibitem[Fearnhead and Rigaill, 2020]{Fearnhead_Rigaill}
Fearnhead, P. and Rigaill, G. (2020).
\newblock Relating and comparing methods for detecting changes in mean.
\newblock {\em To appear in Stat, e291}.

\bibitem[Fryzlewicz, 2014]{Fryzlewicz_WBS}
Fryzlewicz, P. (2014).
\newblock Wild binary segmentation for multiple change-point detection.
\newblock {\em The Annals of Statistics}, 42(6):2243--2281.

\bibitem[{Fryzlewicz}, 2020]{Fryzlewicz_WBS2}
{Fryzlewicz}, P. (2020).
\newblock {Detecting possibly frequent change-points: Wild Binary Segmentation
  2 and steepest-drop model selection}.
\newblock {\em Journal of the Korean Statistical Society}.

\bibitem[Hall et~al., 1990]{HKT90}
Hall, P., Kay, J.~W., and Titterington, D.~M. (1990).
\newblock Asymptotically optimal difference-based estimation of variance in
  nonparametric regression.
\newblock {\em Biometrika}, 77(3):521--528.

\bibitem[Killick et~al., 2012]{Killick_etal}
Killick, R., Fearnhead, P., and Eckley, I.~A. (2012).
\newblock Optimal detection of changepoints with a linear computational cost.
\newblock {\em Journal of the American Statistical Association},
  107(500):1590--1598.

\bibitem[{Kov{\'a}cs} et~al., 2020]{SeedBS}
{Kov{\'a}cs}, S., {Li}, H., {B{\"u}hlmann}, P., and {Munk}, A. (2020).
\newblock {Seeded binary segmentation: a general methodology for fast and
  optimal change point detection}.
\newblock {\em arXiv:2002.06633}.

\bibitem[Kov\'acs et~al., 2020]{OBS}
Kov\'acs, S., Li, H., Haubner, L., B\"uhlmann, P., and Munk, A. (2020).
\newblock {Optimistic search strategies: change point detection without full
  grid search}.
\newblock {\em Working paper}.

\bibitem[Li et~al., 2016]{LMS16}
Li, H., Munk, A., and Sieling, H. (2016).
\newblock F{DR}-control in multiscale change-point segmentation.
\newblock {\em Electronic Journal of Statistics}, 10(1):918--959.

\bibitem[{Londschien} et~al., 2019]{Londschien}
{Londschien}, M., {Kov{\'a}cs}, S., and {B{\"u}hlmann}, P. (2019).
\newblock {Change point detection for graphical models in the presence of
  missing values}.
\newblock {\em To appear in the Journal of Computational and Graphical
  Statistics, arXiv:1907.05409}.

\bibitem[Vostrikova, 1981]{Vostrikova}
Vostrikova, L.~Y. (1981).
\newblock Detecting `disorder' in multidimensional random processes.
\newblock {\em Soviet Mathematics Doklady}, 24:55--59.

\bibitem[Yao, 1988]{Yao1988}
Yao, Y.-C. (1988).
\newblock Estimating the number of change-points via {Schwarz'} criterion.
\newblock {\em Statistics \& Probability Letters}, 6(3):181--189.

\end{thebibliography}

\end{document}